\documentclass[12pt]{article}
 \usepackage{graphicx}
 \usepackage{epsfig}

\usepackage{amssymb}
\usepackage{bm}

\begin{document}


\title{Rise of $Kp$ Total Cross Section and Universality}

\author{Muneyuki Ishida$^a$\thanks{e-mail: mishida@wisc.edu . Address until March 2012 : Department of Physics, University of Wisconsin, Madison, WI 53706} and Vernon Barger$^b$\\
{\small $^a$Department of Physics, School of Science and Engineering, Meisei University, Hino, Tokyo 191-8506, Japan}\\
{\small $^b$Department of Physics, University of Wisconsin, Madison, Wisconsin 53706}
 }

\maketitle

\vspace*{3cm}

\begin{abstract}
The increase of the measured hadronic total cross sections at the highest energies
is empirically described by squared log of center-of-mass energy $\sqrt s$
as $\sigma_{\rm tot}\simeq B$log$^2s$, consistent with the energy dependence of the 
Froissart unitarity bound. 
The coefficient $B$ is argued to have a universal value, but this is not proved directly from QCD. 
In the previous tests of this universality, the  $p(\bar p)p$, $\pi^\mp p$, and $K^\mp p$ forward 
scatterings were analyzed independently and found to be consistent with  
$B_{pp}\simeq B_{\pi p} \simeq B_{Kp}$, although the determined value of $B_{Kp}$ had large uncertainty. 
In the present work, we have further analyzed forward $K^\mp p$ scattering to obtain a more exact value 
of $B_{Kp}$. Making use of continuous moment sum rules(CMSR) we have fully exploited the information of low-energy scattering data to predict the high-energy behavior of the amplitude through duality.    
The estimation of $B_{Kp}$ is improved remarkably, and our result strongly supports the universality of $B$.
\end{abstract}


\section{Introduction and Summary}

The increase of total cross sections $\sigma_{\rm tot}$ in high-energy region
is described\cite{log2} by squared log of center-of-mass energy $\sqrt s$
as $\sigma_{\rm tot}\simeq B$~log$^2s$ consistent with the energy dependence of Froissart unitarity 
bound\cite{froissart}.
The COMPETE Collaboration has further assumed $\sigma_{\rm tot}\simeq B$~log$^2(s/s_0)$
to apply to all hadron total cross sections with a universal value of $B$.\cite{COMPETE}
The universality of $B$ was theoretically anticipated in Refs\cite{jenk,tang} and recently 
inferred from a color glass condensate\cite{cgc,frank} of QCD. However, there is still no rigorous proof
based on QCD.

In the previous work\cite{II}, to test this universality empirically, the  
$p(\bar p)p$, $\pi^\mp p$, and $K^\mp p$ forward scattering amplitudes were analyzed,
and the values of B, denoted respectively as $B_{pp}$, $B_{\pi p}$, and $B_{Kp}$, were estimated independently.
The resulting values were consistent with the universality, 
$B_{pp}\simeq B_{\pi p} \simeq B_{Kp}$, although there was still a rather large uncertainty in $B_{Kp}$,
$B_{Kp}=0.354\pm 0.099$mb. 

In the present work, to reduce the uncertainty of $B_{Kp}$ we have refined the analysis 
of forward $K^\mp p$ scattering. 
By employing continuous moment sum rules(CMSR)\cite{im,dhs,cmsr,barger} for the crossing-even amplitude, 
we have fully exploited the information of low-energy scattering data to predict the high-energy behavior 
of $\sigma_{\rm tot}$ through duality. We use two CMSRs with parameters $\epsilon=-1,-3$ and, 
following ref.\cite{barger}, the estimated values of the cross-section integrals are treated as data points.
The unphysical region of the integral is problematic. The pole of $\Lambda (1405)$ in the unphysical region
 gives a relatively large contribution in $K^-p$ amplitude. 
We estimate its contribution by the coupled-channel study by
 A.~D.~Martin\cite{martin}. The $\sigma_{\rm tot}$ and $\rho$-ratios
($=$Re $f/$Im $f$) of $K^\mp p$ forward amplitudes in high-energy region are fit simultaneously
with two CMSR's data points. 
The resulting value is $B_{Kp}=0.328\pm 0.045$mb. 
The error is improved remarkably, less than half of the previous estimate. 
By comparing with previously determined values of $B_{pp}$ and $B_{\pi p}$,
the universality, $B_{Kp}\simeq B_{\pi p} \simeq B_{pp}$, is strongly supported.

\section{Formulas}

We inherit the notation and definitions from the previous work\cite{II}:
$\nu (k)$ is the energy(momentum) of the kaon beam in laboratory system, $\nu =\sqrt{k^2+m_K^2}$
where $m_K$ is mass of $K^\mp$. It is related to the center of mass energy $\sqrt s$ by
\begin{eqnarray}
s &=& M^2+m_K^2+2M\nu\ \ .
\label{eq1}
\end{eqnarray}
The $K^\mp p$ forward amplitudes are denoted as $f^{K^\mp p}(\nu )$.
The total cross sections $\sigma_{\rm tot}^{K^\mp p}$ are given by Im~$f^{K^\mp p}(\nu )=(k/4\pi)\sigma_{\rm tot}^{K^\mp p}$ through optical theorem.  
The crossing-even$/$odd amplitudes $f^{(\pm )}(\nu )$ are given by 
$f^{(\pm )}(\nu )=( f^{K^-p}(\nu )\pm f^{K^+p}(\nu) )/2$.
The $f^{(+)}(\nu )$ is related to the $A^\prime (\nu ,t)$ amplitude at $t=0$\cite{barger} as
$f^{(+)}(\nu )= A^\prime (\nu ,t=0)/4\pi$.
Real and imaginary parts of $f^{(\pm )}(\nu )$ are assumed to take the forms\cite{II}
\begin{eqnarray}
{\rm Im}~f_{\rm as}^{(+)} &=& \frac{\nu}{m_K^2}\left( c_2{\rm log}^2\frac{\nu}{m_K} +c_1{\rm log}\frac{\nu}{m_K}+c_0 \right)
+\frac{\beta_{P^\prime}}{m_K}\left(\frac{\nu}{m_K}\right)^{\alpha_{P^\prime}}\label{eq2}\\
{\rm Im}~f_{\rm as}^{(-)} &=& \frac{\beta_{V}}{m_K}\left(\frac{\nu}{m_K}\right)^{\alpha_{V}}\label{eq3}\\
{\rm Re}~f_{\rm as}^{(+)} &=& \frac{\pi\nu}{2m_K^2}\left( 2c_2{\rm log}\frac{\nu}{m_K} +c_1 \right)
-\frac{\beta_{P^\prime}}{m_K}\left(\frac{\nu}{m_K}\right)^{\alpha_{P^\prime}}
{\rm cot}\frac{\pi\alpha_{P^\prime}}{2} + f^{(+)}(0)\label{eq4}\\
{\rm Re}~f_{\rm as}^{(-)} &=&\frac{\beta_{V}}{m_K}\left(\frac{\nu}{m_K}\right)^{\alpha_{V}}
{\rm tan}\frac{\pi\alpha_{V}}{2} \ \ ,\label{eq5}
\end{eqnarray}
in the asymptotic high-energy region,
where we assume Im~$f^{(+)}$ is given by log$^2\nu$(and log~$\nu$) terms in addition to the ordinary
Pomeron($c_0$) term and Reggeon($\beta_{P^\prime}$) term. Similarly Im~$f^{(-)}$ is given by
exchange of $\rho ,\omega$-meson trajectories in Regge theory with degenerate trajectory intercepts assumed.
Their intercepts are taken to be an empirical value
$\alpha_{P^\prime}\simeq \alpha_{V}\simeq 0.5$ .
The real parts of $f_{\rm as}^{(\pm )}$ are obtained from their imaginary parts by using crossing-symmetry,
$f^{(\pm )}(-\nu_R -i\epsilon)=\pm f^{(\pm)}(\nu_R+i\epsilon )$, except for a subtraction constant $f^{(+)}(0)$.
%
There are the total six parameters: $c_{2,1,0}$, $\beta_{P^\prime ,V}$, $f^{(+)}(0)$. 
The $c_2$-term in Im~$f_{\rm as}^{(+)}$ dominates $\sigma_{\rm tot}$ in the 
high-energy region and it is related to $B_{Kp}$ by
\begin{eqnarray}
B_{Kp} &=& 4\pi c_2/m_K^2  \ \ .
\label{eq6}
\end{eqnarray}

\section{Continuous-Moment Sum Rules}

We can exploit the low-energy scattering data to predict the amplitudes in high-energy region
by using continuous-moment sum rules(CMSR)\cite{cmsr}. 
In ref.\cite{barger} a compact form of CMSR is given in $\pi N$ scattering,
where Regge-pole contributions are parametrized in a form satisfying crossing-symmetry
and convenient for CMSR.
The crossing-even $A^\prime (\nu ,t)$ amplitude was taken to be 
$A^\prime (\nu ,t )= \sum_{i=P,P^\prime ,P^{\prime\prime}}[ -\gamma_i(\nu_0^2-\nu^2)^{\alpha_i/2} ]$
in the asymptotic energy region $\nu\ge \nu_1$ . 
Here $\nu_0$ is the normal threshold $\nu_0= \mu + t/(4M)$, where $\mu (M)$ is charged pion(proton) mass, and 
$(\nu_0^2-\nu^2)$ is analytically continued to 
$(\nu^2-\nu_0^2)e^{-i\pi}$ in $\nu > \nu_0$ for $A^\prime (\nu ,t)$ to have the correct phase factor of 
Reggeon-exchange amplitudes.
The finite-energy sum rule for $\nu A^\prime (\nu ,t)$ is given by
\begin{eqnarray}
&&\int_{\nu_0}^{\nu_1} d\nu\  \nu {\rm Im}[(\nu_0^2-\nu^2)^{-(\epsilon +1)/2} A^\prime] 
= \sum_{i=P,P^\prime ,P^{\prime\prime}} \gamma_i 
\frac{(\nu_1^2-\nu_0^2)^{(\alpha_i-\epsilon+1)/2}{\rm sin}
[\frac{\pi}{2}(\alpha_i-\epsilon-1)]}{\alpha_i-\epsilon+1}\ \ 
\label{eq7}
\end{eqnarray}
where $\epsilon$ is a continuous parameter and the nucleon pole term contribution should be added 
to the left hand side(LHS).

If we know completely the scattering amplitudes in low-energy region from experiments, 
the high-energy amplitudes are predicted via CMSR through analyticity. 
In ref.\cite{barger}, by using the CERN results on $\pi N$ scatterings up to 2~GeV, 
the low energy integrals on the LHS of Eq.~(\ref{eq7}) were evaluated in the region 
$0\le -t < 1$~GeV$^2$ and $0(-1)\ge \epsilon \ge -5$ for $t=0(t<0)$ in steps of 0.5 .
These numbers were treated as data points, and simultaneously fit as data to the Reggeon
parametrization of the asymptotic high-energy region. 

This method can be applied to the analysis of forward $K^\mp p$ scattering. 
Our $f^{(+)}(\nu )$ amplitude corresponds to $A^\prime /(4\pi )$ with $t=0$, and the CMSR is given by
\begin{eqnarray}
&&\int_{0}^{\nu_1}d\nu\ \nu {\rm Im}~[(m_K^2-\nu^2)^{-(\epsilon+1)/2}f^{(+)}(\nu)]
= \int_{0}^{\nu_1}d\nu\ \nu {\rm Im}~[(m_K^2-\nu^2)^{-(\epsilon+1)/2}f_{\rm as}^{(+)}(\nu)].\ \ 
\label{eq8}
\end{eqnarray}
The LHS should be evaluated from low-energy experimental data, while the RHS is
analytically calculated by using the formulas (\ref{eq2}) and (\ref{eq4}).

If we take non-odd values of $\epsilon$, Re~$f^{(+)}(\nu )$ data in low-energy region are necessary as inputs.
However, experimental data\cite{pdg} of Re~$f(\nu )$(or $\rho$ ratios) for $K^\mp p$ are poorly known 
for $k\le 5$~GeV. 
In this situation we are forced to select $\epsilon$ as odd integers; specifically we take $\epsilon =-1,-3$.

The CMSRs with $\epsilon =-1,-3$ are equivalent to the $n=1,3$ moment sum rule\cite{dhs},
\begin{eqnarray}
\frac{2}{\pi}\int_0^{\nu_1}d\nu\ \nu^n{\rm Im}~f^{(+)}(\nu)
&=& \frac{2}{\pi}\int_0^{\nu_1}d\nu \  \nu^n{\rm Im}~f_{\rm as}^{(+)}(\nu ),\ \ 
\label{eq9}
\end{eqnarray}
where 
$\frac{2}{\pi}$ is from our convention.
The LHS of Eq.~(\ref{eq9}) is evaluated in the next section.

\section{Evaluation of integrals from experimental data}

The LHS of Eq.~(\ref{eq9}) is obtained by averaging the integrals of Im~$f^{K^+p}$ and
Im~$f^{K^-p}$, which are evaluated separately from experimental data. 

The $K^+p$ channel is exotic and it has no contribution below threshold, $\nu < m_K$. Thus, 
its integral region is $m_K$ to $\nu_1$. By changing the variable from $\nu$ to $k$,
the relevant integral is given by
\begin{eqnarray} 
\frac{2}{\pi}\int_0^{\nu_1}d\nu\ \nu^n{\rm Im}~f^{K^+p} &=&
 \frac{2}{\pi}\int_0^{\overline{\nu_1}}dk\ k\nu^{n-1}\frac{k}{4\pi}\sigma_{\rm tot}^{K^+p}\ \ \ \ 
\label{eq10}
\end{eqnarray}
where $\overline{\nu_1}\equiv \sqrt{\nu_1^2-m_K^2}$, which is taken to be some value in 
asymptotic high-energy region. We take $\overline{\nu_1}=5$~GeV.
Actually the integral of RHS of Eq.~(\ref{eq10}) is estimated by dividing its region into two parts: 
In the low-energy part, from $k=0$ to $k_d$, there are many data points\cite{pdg}. 
They are connected by straight lines and the area of this polygonal line graph can be regarded as the 
relevant integral. In high-energy part, from $k_d$ to $\overline{\nu_1}$, we use the phenomenological fit
used in our previous work\cite{II}. The dividing momentum $k_d$ is taken to be 3~GeV. As a result we obtain 
\begin{eqnarray} 
&& \frac{2}{\pi}\int_0^{\nu_1}d\nu\ \nu^n{\rm Im}~f^{K^+p} = \frac{2}{\pi}\int_{0}^{k_d}dk\cdots
 + \frac{2}{\pi}\int_{k_d}^{\overline{\nu_1}}dk\ \cdots 
 \nonumber\\
&&= 
\left\{ \begin{array}{l}
20.348(41) + 72.435(184) = 692.795(189){\rm GeV} \\
115.15(27) + 1294.90(3.48) = 1410.05(3.49){\rm GeV}^3\ \ \ \  \\
\end{array}\right.    \ 
\label{eq11}
\end{eqnarray}
for $n=1,3$ respectively.

On the other hand, $K^-p$ amplitude includes Born terms of $\Sigma^0,\Lambda^0$ poles, which correspond to the 
nucleon pole term in $\pi N$ amplitude. $K^-p$ is an exothermic reaction with open channels,
$\Lambda \pi$ and $\Sigma \pi$, below threshold which give contributions to Im~$f^{K^-p}(\nu )$ 
in unphysical regions, $\nu_{\Lambda\pi} \le \nu \le m_K$ and  $\nu_{\Sigma\pi} \le \nu \le m_K$, respectively,
where $\nu_{\Lambda\pi}/\nu_{\Sigma\pi}$ is $\Lambda\pi/\Sigma\pi$ threshold energy,
$\nu_{\Lambda\pi/\Sigma\pi}=[(M_{\Lambda^0/\Sigma^0}+\mu)^2-M^2-m_K^2]/(2M)$.
A large contribution from $I=0$ $\Lambda (1405)$ is expected to be in the unphysical region.  
In order to estimate these contributions we adopt a coupled-channel analysis by 
A.~D.~Martin\cite{martin}. 
In his analysis the cross sections of $K^-p\rightarrow K^-p,\bar K^0n,\Sigma\pi ,\Lambda\pi^0$;\  
$K_2^0p\rightarrow K_1^0p,\Lambda\pi^+$; \ $\sigma_{\rm tot}(K_2^0p)$, and Re~$f^{K^-p,K^-n}$
were reproduced successfully. 
Martin included a correction from Coulomb scattering following the scheme of 
Dalitz and Tuan\cite{tuan} for channels like $K^-p$,
which gives a fairly large effect in very low energy region. 
Here we use the purely strong-interaction part of Martin's amplitude
to estimate the unphysical region contribution.
We consider this is the most reliable way to estimate the unphysical region.
The relevant channels are the three $I=1$ channels $\bar KN,\Sigma\pi,\Lambda\pi$ and 
the two $I=0$ channels $\bar KN,\Sigma\pi$.
The $S$-wave $I=1(0)$ scattering amplitudes ${\bm T}^{I=1}({\bm T}^{I=0})$ are parametrized
by the $K$-matirix ($M$-matrix) as
${\bm T}^{I=1}={\bm K}(1-i{\bm q}{\bm K})^{-1}$ (${\bm T}^{I=0}=({\bm M}-i{\bm q})^{-1}$),
where ${\bm q}$ is a diagonal matrix of the channel c.m. momenta denoted as 
${\bm q}=diag\{ p,p_\Sigma ,p_\Lambda \}(diag\{ p,p_\Sigma \})$.
The real symmetric 3-by-3 matrix ${\bm K}$ is taken to be constant, including six parameters.
The ${\bm M}$ matrix is taken to be effective range form, ${\bm M}={\bm A}+{\bm R}p^2$, 
which makes it possible to describe $\Lambda (1405)$ resonance.
Here $p$ is the $\bar KN$ c.m. momentum 
which is continued below threshold as $p=i|p|$. 
Real symmteric 2-by-2 matrices ${\bm A},{\bm R}$ are taken to be constant, 
and ${\bm M}$ includes six parameters. 

By using the elements of ${\bm K}$ and ${\bm M}$, 
the invese of I=1,0 $\bar KN$ scattering amplitudes ${\bm T}^{I=1,0}_{KK}$
are given explicitly\cite{ross} by 
\begin{eqnarray}
({\bm T}_{KK}^{I=1,0})^{-1} &=& A_{I=1,0}^{-1} + ip \nonumber\\
A_{I=1} &=&  K_{KK} + \frac{1}{B_1}(K_{K\Sigma}B_2+K_{K\Lambda}B_3)\nonumber\\
   B_1 &=&(1-ip_\Sigma K_{\Sigma\Sigma})(1-ip_\Lambda K_{\Lambda\Lambda}) + p_\Sigma p_\Lambda K_{\Sigma\Lambda}^2\nonumber\\
   B_2 &=&p_\Sigma p_\Lambda (K_{K\Sigma}K_{\Lambda\Lambda}-K_{K\Lambda}K_{\Sigma\Lambda}) 
     +i p_\Sigma K_{K\Sigma} \nonumber\\
   B_3 &=&p_\Sigma p_\Lambda (K_{K\Lambda}K_{\Sigma\Sigma}-K_{K\Sigma}K_{\Sigma\Lambda}) 
     +i p_\Lambda K_{K\Lambda}
\nonumber\\
A_{I=0}^{-1} &=& M_{KK} - \frac{M_{K\Sigma}^2}{M_{\Sigma\Sigma}-ip_\Sigma} 
\label{eq11c}
\end{eqnarray}
where 
the subscript $_{KK}$ means (1,1) element corresponding to $\bar KN\rightarrow \bar KN$. 
Other subscripts are used similarly.
$A_{I}=a_I+ib_I$ are the $S$-wave $\bar KN$ scattering lengths and $M_{KK}=A_{KK}+R_{KK}p^2$ etc.
The best-fit values of the relevant 12 parameters are given\cite{comment1} with errors in ref.\cite{martin}.

Our $K^-p$ forward amplitude $f^{K^-p}(\nu )$ is 
related\cite{comment2} to the ${\bm T}_{KK}^I$ by
\begin{eqnarray}
f^{K^-p}(\nu )=\frac{\sqrt s}{2M}[ {\bm T}^{I=0}_{KK}+{\bm T}^{I=1}_{KK} ]\ .
\label{eq12}
\end{eqnarray}
There are no $\sigma_{\rm tot}^{K^-p}$-data reported below $k\equiv k_s=0.245$~GeV in Particle Data Group\cite{pdg}.
The Im~$f^{K^-p}$ obtained by Martin is also utilized in this energy-region $m_K \le \nu \le \nu_s$ where
$\nu_s=\sqrt{k_s^2+m_K^2}$.

The relevant integral is evaluated as follows:\\
\begin{eqnarray}
&&\frac{2}{\pi}\int_0^{\nu_1}\nu^n{\rm Im}~f^{K^-p}(\nu )d\nu \nonumber\\
&&= 
\sum_{R=\Lambda^0,\Sigma^0}\frac{g_R^2}{M} \nu_{B,R}^n(-M_R+M+\nu_{B,R})
+\frac{2}{\pi}\int_{\nu_{\Lambda\pi}}^{\nu_s}d\nu\cdots + \frac{2}{\pi}\int_{k_s}^{k_d}dk\cdots 
+\frac{2}{\pi}\int_{k_d}^{\overline{\nu_1}}dk\cdots 
\nonumber\\
&&= \left\{
\begin{array}{l} (-0.106\pm 0.060) + (0.508\pm 0.095) + (35.481\pm 0.069)+(108.499\pm 0.405) \\
(-0.0004\pm 0.0010) + (0.110\pm 0.017) + (191.28\pm0.54) + (1928.04\pm 7.46) \\ 
\end{array}\right.
\label{eq13}\\
&&=\left\{
\begin{array}{lc} 
 144.382\pm 0.426~{\rm GeV} & n=1\\  2119.43\pm 7.48\ {\rm GeV}^3 & n=3 \\
\end{array}\right.
\label{eq14}
\end{eqnarray}
where $\nu_{B,R}=(M_R^2-M^2-m_K^2)/(2M)$.
The 1st bracket of Eq.~(\ref{eq13}) is the Born-term, which is estimated by using 
$g_{\Lambda^0}^2=13.7\pm 1.9$ and $g_{\Sigma^0}^2=0$. The error comes from the upper limit of $g_{\Sigma^0}^2<3.7\pm 1.3$\cite{martin}. 
The 2nd bracket, corresponding to the integral region $\nu_{\Lambda\pi}\le\nu\le\nu_s$, is also estimated by 
Eq.~(\ref{eq12}) of Martin's amplitude. The error comes from the $R_{KK}=0.41\pm 0.10$~fm\cite{martin},
which gives the largest uncertainty among all the parameters.
The 3rd and 4th brackets are estimated from the polygonal line graph and phenomenological fit\cite{II}, 
respectively.
For $n=1$ a small but sizable contribution comes from 1st term and 2nd term.
The corresponding errors affect the final value of 
Eq.~(\ref{eq14}) but they are not main sources of its error.
For $n=3$ the 1st and 2nd terms are both negligible. 
Thus, the uncertainty from the unphysical region, especially from $\Lambda (1405)$ pole,
is considered to be insignificant in use of the values of Eq.~(\ref{eq14}).

By averaging Eqs.~(\ref{eq11}) and (\ref{eq14}) we obtain
\begin{eqnarray}
\frac{2}{\pi}\int_0^{\nu_1}\nu^n{\rm Im}~f^{(+)}(\nu )d\nu 
&=& \left\{ \begin{array}{l}
118.589 \pm 0.233\ {\rm GeV} \ \\
1764.74 \pm 4.13\ {\rm GeV}^3 \ 
\end{array}\right.\ \ \ \ 
\label{eq15}
\end{eqnarray}
for $n=1,3$ respectively.
These two values are treated as low-energy data points and they are fit simultaneously with the data 
in asymptotic high-energy region.

\begin{figure}
\begin{center}
\includegraphics{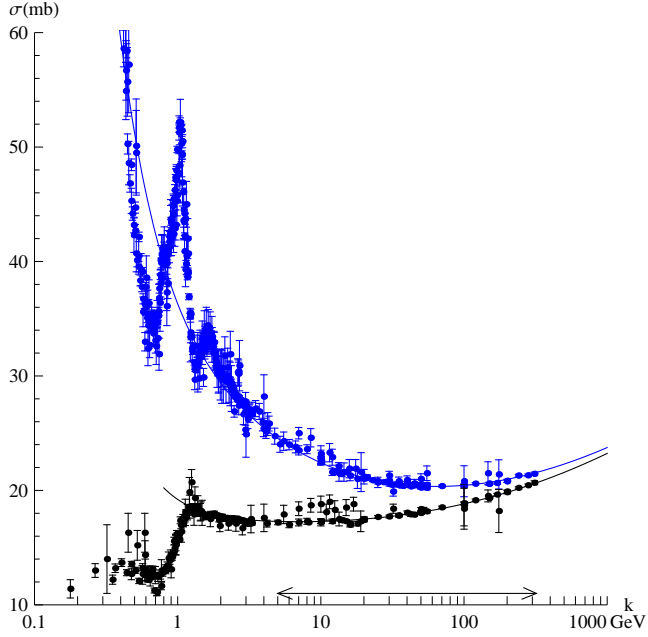}
\end{center}
\caption{\label{fig1} Results of the fit to $\sigma_{\rm tot}^{K^\mp p}$(mb). Upper(blue) data represent $K^-p$
and lower(black) data are $K^+p$. The horizontal arrow represents the energy region
of the fit to the asymptotic amplitude. The solid curves are the asymptotic Reggeon amplitudes,
which are extrapolated down through the resonance region.}
\end{figure}

\begin{figure}
\begin{center}
\includegraphics{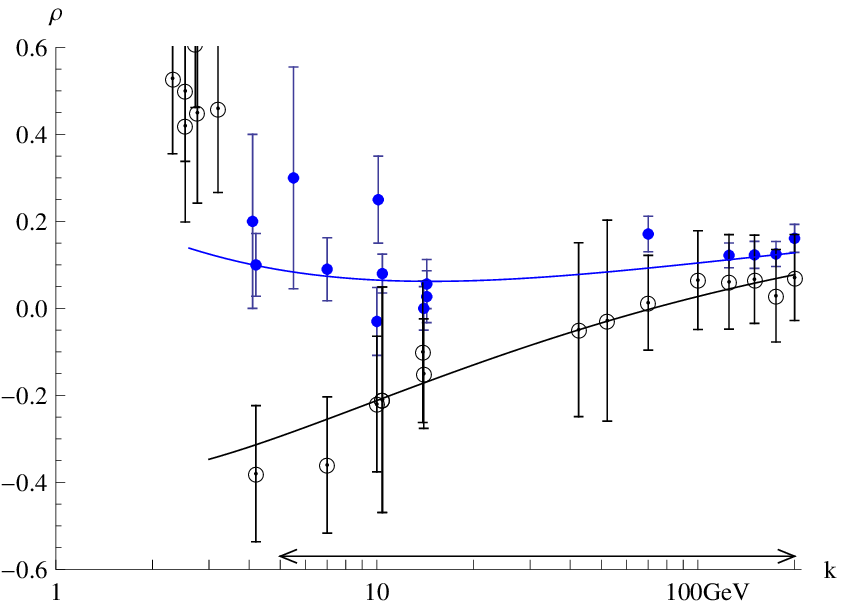}
\end{center}
\caption{\label{fig2} Results of the fit to $\rho^{K^\mp p}$. The solid(blue) data points represent $K^-p$
and the open(black) data points are $K^+p$. The horizontal arrow represents energy region
of the fit to the asymptotic amplitude.}
\end{figure}

\section{Results and Concluding Remarks}
The $\sigma_{\rm tot}^{K^-p}$, $\sigma_{\rm tot}^{K^+p}$, $\rho^{K^-p}$, and $\rho^{K^+p}$
(more precisely\cite{comment3} Re~$f^{K^-p}$, and Re~$f^{K^+p}$
)
with $k\ge 5$~GeV, which are given in ref.\cite{pdg}, 
are fit by using the formula in \S 2. 
The numbers of parameters are six: $c_{2,1,0}$, $\beta_{P^\prime ,V}$, $f^{(+)}(0)$.
We fit to the CMSR data points of $n=1,3$ (\ref{eq15}) simultaneously. We considered three cases:
i) the case including a $n=1$ datum of (\ref{eq15}), 
ii) the case including a $n=3$ datum of (\ref{eq15}), and
iii) the case including both $n=1,3$ data of (\ref{eq15}).
The results are compared with iv) the case with no use of CMSR and also 
our previous analysis II\cite{II}.

The best fit parameters and $\chi^2$ values are given in Table \ref{tab1}.

\begin{table*}
\caption{\label{tab1} Best fit parameters and $\chi^2$: 
$\sigma_{\rm tot}^{K^\mp p}$ and Re~$f^{K^\mp p}$ in $k\ge 5$~GeV are fit simulaneouly with 
i) $n=1$ CMSR datum, ii) $n=3$ CMSR datum, iii) $n=1$ and 3 CMSR data.
The results are compared with the case iv) with no CMSR data and with the previous analysis\cite{II},
 where $\sigma_{\rm tot}^{K^\mp p}$ in $k\ge 20$~GeV and Re~$f^{K^\mp p}$ in $k\ge 5$~GeV
are fit using the finite-energy sum rule with region of the integral 5$\le$$k$$\le$20~GeV as a constraint.   
}
\begin{tabular}{l|cccccc|l}
case & $c_2$ & $c_1$ & $c_0$ & $\beta_{P^\prime}$ & $\beta_V$ & $f^{(+)}(0)$ & $\chi^2/(N_D-N_P-1)$\\
\hline  
i) $n=1$ &  0.01652(224) & -0.1241 & 1.152 & 0.2702 & 0.5741 & 1.180 & $143.21/$(165-6-1)\\
ii) $n=3$ & 0.01724(256) & -0.1339 & 1.188 & 0.2110 & 0.5736 & 1.609 & $143.54/$(165-6-1)\\
iii) $n=1,3$ both & 0.01634(223) & -0.1221 & 1.146 & 0.2736 & 0.5737 & 1.178 & $144.05/$(166-6-1)\\
iv) no CMSR & 0.01522(385) & -0.1065 & 1.088 & 0.3726 & 0.5749 & 2.104 & $143.04/$(164-6-1)\\
\hline
II\cite{II} & 0.01757(495) & -0.1388 & 1.207 & 0.1840 & 0.5684 & 1.660 & $63.80/$(111-5-1)\\
\end{tabular}
\end{table*}
All fits are successful ( $\chi^2/$deg.freedom$<1$ ).  
The best-fit $\chi^2$ values of i),ii),iii) are almost the same as the case iv) with no use of CMSR, 
suggesting that the CMSR works well in the fit.
Results of the best fit in the case iii) are shown in Fig.~\ref{fig1} and \ref{fig2}.

The error of $c_2$ in the case iii) is much smaller than the case iv), and is greatly improved from our previous
analysis II\cite{II}.
Correspondingly the parameter $B_{Kp}$ associated by the ln$^2s$ dependence is given by
\begin{eqnarray}
\begin{array}{lcl}
{\rm Present\ result\ iii)} & & {\rm Previous\ analysis\ II}\cite{II}\\
c_2=0.01634\pm 0.00223 & \leftarrow & 0.01757\pm 0.00495 \\
B_{Kp}=0.328\pm 0.045\ {\rm mb} & \leftarrow & 0.354\pm 0.099\ {\rm mb}\\ 
\end{array}
\label{eq16}
\end{eqnarray}

There are several comments should be added:\\
i) 
In the present analysis we take $\overline{\nu_1}=5$~GeV, and the $\sigma_{\rm tot}$ and Re~$f$ data 
in $k\ge \overline{\nu_1}$ are fit. The results are almost independent of the choice of 
$\overline{\nu_1}$. If we take $\overline{\nu_1}=4$~GeV and the CMSR with $\epsilon =-1,-3$ 
together with the data of $k\ge 4$~GeV are fit simultaneously, we obtain the result 
$B_{Kp}=0.311\pm 0.039$~mb which is almost the same result as Eq.~(\ref{eq16}).
\\
ii) 
In our previous analysis\cite{II}, $\overline{\nu_1}=20$~GeV was taken, and the 
the data of $\sigma$ in $k\ge 20$~GeV and Re~$f(k)$ in $k\ge 5$~GeV were fit. 
The $n=1$ integral with $\overline{\nu_1}=20$~GeV of LHS (\ref{eq11})(, which 
is obtained as 6802.61(10.90)~GeV) 
is fit simultaneously to the data of the same energy-regions as the previous 
analysis\cite{II}. The resulting value is $B_{Kp}=0.360\pm 0.110\ {\rm mb}$, which is almost 
the same as II\cite{II},  $B_{Kp}=0.354\pm 0.099$~mb so no improvement is obtained in this method. 
Hence we have adopted a different energy-region in the present analysis.

\begin{figure}
\begin{center}
\includegraphics{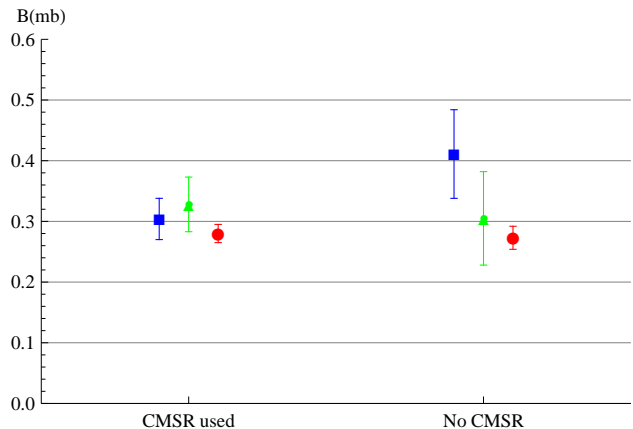}
\end{center}
\caption{\label{fig3} Values of the asymptotic parameter $B$(mb) for the $p(\bar p)p,\ \pi p$, and $Kp$ 
total cross sections using CMSR(left) in comparison with the cases with no CMSR constraint(right). 
Red circles and blue squares represent the 
$p(\bar p)p$ and $\pi p$, respectively, which were obtained in our previous analysis II\cite{II}. 
Green triangles are $Kp$, obtained in the present analysis.}
\end{figure}

The obtained value of $B_{Kp}$ together with the previous estimates of $B_{\pi p}$ and $B_{pp}$ in ref.\cite{II}
are shown graphically in Fig.~\ref{fig3} in compared with the case with no use of sum rules.
Our results strongly suggest the universalty of the coefficient $B$. 
It should be noted that the inclusion of low-energy data through the CMSR is essential 
in reaching this conclusion,
especially that $B_{Kp}\simeq B_{\pi p}$ as shown in the present work. 
Our conclusion is that the $B$ does not depend on the flavor content of the particles scattered.

\vspace*{1cm}


\noindent{\it Acknowledgements}

M.I. is very grateful to Professor M.~G.~Olsson, Professor F.~Halzen for helpful comments 
and discussions.
He also thanks the members of phenomenology institute of University of Wisconsin-Madison
for hospitalities. He expresses his sincere gratitude to Professor K.~Igi for his thoughtful suggestions and encouragement.
This work was supported in part by the U.S. Department of Energy under grant No. DE-FG02-95ER40896,
in part by KAKENHI(2274015, Grant-in-Aid for Young Scientists (B)) and 
in part by grant as Special Researcher of Meisei University. 

\nocite{*}

\bibliography{apssamp}

\begin{thebibliography}{00}
\bibitem{comment1} The best-fit values\cite{martin} are 
 $$ \begin{array}{ccc} { K}_{KK}=1.01(7) & { K}_{K\Sigma}=-0.92(5) &{ K}_{K\Lambda}=-0.66(5)\\
 { K}_{\Sigma\Sigma}=0.58(19) & { K}_{\Sigma\Lambda}=0.62(24) &{ K}_{\Lambda\Lambda}=0.08(25)\\
 { A}_{KK}=-0.10(5) & { A}_{K\Sigma}=-0.92(3) &{ A}_{\Sigma\Sigma}=1.60(16)\\
 { R}_{KK}=0.41(10) & { R}_{K\Sigma}=-0.30(18) & { R}_{\Sigma\Sigma}=0.07(35) \\
  \end{array} $$ 
where the elements of ${\bm K},{\bm R}({\bm A})$ are given in fm(fm$^{-1}$).
\bibitem{comment2} The factor $\frac{\sqrt s}{M}$ comes from different normalizations of $f$ and ${\bm T}$,
Im~$f=\frac{k}{4\pi}\sigma$ and Im~$({\bm T})_{KK}=\frac{p}{4\pi}\sigma$. 
The laboratory momentum $k$ is related to the c.m. momentum $p$ by $k=\frac{\sqrt s}{M}p$.
\bibitem{comment3} In the analysis we obtained Re~$f^{K^\mp p}$ data points from
original $\rho^{K^\mp p}$ data\cite{pdg} by multiplying by 
Im~$f^{K^\mp p}=\frac{k}{4\pi}\sigma_{\rm tot}^{K^\mp p}(k)$, where we use as $\sigma_{\rm tot}^{K^\mp p}(k)$
the fit result of ref.\cite{pdg}. See, ref.\cite{II} for more detail. 
In the analysis 
$\sigma_{\rm tot}^{K^\mp p}$ data and  Re~$f^{K^\mp p}$ data are fit simultaneously.
\end{thebibliography}

\end{document}